\documentclass[conference]{IEEEtran}
\usepackage{cite}
\usepackage{amsmath,amssymb,amsfonts}
\usepackage{graphicx}
\usepackage{textcomp}
\usepackage{booktabs}
\usepackage{array}
\usepackage{xcolor}
\usepackage[table]{xcolor}
\newcommand{\Tset}{\mathcal T}
\newcommand{\J}{\mathcal J}
\newcommand{\Sset}{\mathcal S}
\newcommand{\F}{\mathcal F}
\newcommand{\RT}{\mathrm{RT}}
\newcommand{\DA}{\mathrm{DA}}
\newcommand{\HP}{\mathrm{HP}}

\begin{document}

\title{\vspace{-0.88\baselineskip} Solver-Free and Cost-Aware Online Scheduling \\ of AI Data Center Tasks}
\IEEEaftertitletext{\vspace{-2.88\baselineskip}}

\author{\IEEEauthorblockN{Xin Qin, Dawei Qiu}
\IEEEauthorblockA{\textit{School of Electrical and Electronic Engineering} \\
\textit{Nanyang Technological University}, Singapore \\
\{xin.qin, dawei.qiu\}@ntu.edu.sg}
\and
\IEEEauthorblockN{Fei Teng}
\IEEEauthorblockA{\textit{Department of Electrical and Electronic Engineering} \\
\textit{Imperial College London}, London, UK \\
f.teng@imperial.ac.uk}
}

\maketitle

\begin{abstract}
AI data centers receive computing tasks continuously, whereas solver-based optimization takes time to run and electricity is priced per settlement period. This paper bridges this timing gap through a solver-free online optimization rule inspired by dual decomposition. The rule assigns each task a start period upon arrival to optimize energy cost subject to operational constraints, shifting flexible tasks toward low-price periods while keeping high-priority tasks on schedule. Simulations on the A100 GPU trace (3,456 GPUs, 1,343 tasks) show that the proposed rule keeps GPU and CPU use within capacity while scheduling tasks at low cost. Compared with a multi-timescale optimization benchmark, it achieves a lower cost and, because it needs no solver, much less computation time.
\end{abstract}

\vspace{-0.28em}

\section{Introduction}
\vspace{-0.28em}
\label{sec:intro}
\subsection{Background}
\vspace{-0.28em}
\label{sec:intro.A}
AI data centers (AIDCs), which host training and inference clusters, are now large electrical loads, and electricity has become their largest operating expense, accounting for about two-thirds of the operating cost~\cite{epoch2026}. Their cost is also increasingly exposed to wholesale prices, which vary within the day.
To reduce energy cost, part of their workload can be shifted to low-price periods by delays of minutes to hours, and experiments on GPU clusters have shown power reductions without compromising service targets~\cite{colangelo2026,williams2026}. These findings motivate scheduling the start times of delay-tolerant tasks to exploit lower-price periods while still completing every task.

The difficulty is a mismatch between the timescales of computing and energy. On the computing side, tasks arrive continuously, time-stamped to the second~\cite{duan2026gfs}. On the energy side, electricity is bought day ahead or in real time priced per hour or per 15~min, and an energy optimization that decides task start periods requires computational resources and time. This creates a dilemma: If the AIDC operator solves the optimization whenever a task arrives, the solver may not finish before the next arrival; If the operator solves it once per period instead, tasks wait unnecessarily for the next solve and may miss low-price periods.

This paper therefore address one question: Can a solver-free rule that schedules each task once, on arrival, respect the GPU and CPU capacities while optimizing energy cost?

\subsection{Literature Review}
\vspace{-0.18em}
\label{sec:intro.B}
Existing works on energy-aware task scheduling in data centers follow three approaches.

First, horizon-based methods plan a given workload over a given horizon. For example, the method in~\cite{radovanovic2023} uses day-ahead carbon forecasts to cap the next day’s hourly capacity for flexible workloads. Recent AIDC studies schedule training and inference workloads by robust optimization under uncertainty~\cite{yang2026} or by a MILP for an AIDC operating as a microgrid prosumer~\cite{zhang2026}. These methods suit advance scheduling, when all tasks and prices are known, but are difficult to apply in real time as tasks arrive.

Second, online methods decide each job as it arrives. Paper~\cite{hanafy2024} delays flexible tasks into the lowest-carbon period within a waiting limit, also considering cloud service cost. Work~\cite{he2022} schedules tasks slot by slot under stochastic prices, and work~\cite{wang2014} proposes a randomized online job and energy optimizer with a competitive ratio, building on online allocation theory~\cite{buchbinder2009,azar1998}. These rules target carbon, service cost or worst-case guarantees, rather than the energy cost minimization under resource limits.

Third, multi-timescale methods separate fast and slow decisions. Work~\cite{yao2012} switches servers on a slow time scale and sets service rates every slot, with Lyapunov bounds on cost and delay. Work~\cite{hall2025} shapes the load one day ahead and places arriving jobs in real time, and D3T~\cite{ran2026} assigns queued tasks to different sites at a time scale of seconds and optimizes cooling every 30~min. These methods decide where a task runs or how fast queued work is served, which may rely on offline training, however, they do not decide when flexible tasks start to reduce energy costs.

\subsection{Contributions}
\vspace{-0.18em}
\label{sec:intro.C}
This paper proposes an online optimization method that schedules AIDC tasks in real time. We take the perspective of the AIDC operator, as a price-taker, and schedule the tasks submitted during an operating day as they arrive. The contributions are summarized below:

\textit{1) A fast online scheduling method.} The proposed method decides each task once, at submission, by choosing its start period within the allowed delay, and achieves high computational efficiency without a solver.

\textit{2) A cost-efficient rule based on dual decomposition.} Inspired by dual decomposition, the proposed method integrates the CPU/GPU resource shadow prices and reserves capacity, which contributes to cost optimization and keeps high-priority tasks on time. 

\textit{3) Case studies with benchmark comparisons.} Based on the practical AI task data in the A100 GPU trace, we investigate the operational cost and computation time of the proposed method by comparing it with an energy-unaware rule, a multi-timescale optimization, and the perfect offline optimum.

The remainder of this paper is organized as follows. Section~\ref{sec:model} introduces the problem setting and formulation. Section~\ref{sec:method} proposes the online rule. Section~\ref{sec:results} presents case studies and comparisons with benchmarks, and Section~\ref{sec:conclusion} presents conclusions and discussion.

\section{Model Formulation}
\label{sec:model}
We introduce the model formulation in this section. For clarity, we list the notation in Table~\ref{tab:nom}.

\begin{table}[!t]
\caption{Nomenclature}
\vspace{-0.88em}
\label{tab:nom}
\centering
\footnotesize
\setlength{\tabcolsep}{3pt}
\begin{tabular}{@{}l>{\raggedright\arraybackslash}p{0.74\columnwidth}@{}}
\toprule
Symbol & Meaning\\
\midrule
$t\in\Tset$ & Time period, $\Tset=\{1,\ldots,T\}$\\
$c\in\{G,C\}$ & Resource: GPU or vCPU\\
$j\in\J$, $\J^{\HP}$ & Task submitted; set of all and high-priority (HP) tasks\\
$\Sset_j$ & Set of possible start periods of task $j$\\
$\F_j$ & Set of feasible start periods of task $j$ (fit free capacity)\\
\midrule
$\Delta t$ & Operation period length\\
$\lambda_t^{\DA}$, $\lambda_t^{\RT}$ & Day-ahead and real-time price in period $t$ (\$/MWh)\\
$s_j$, $\ell_j$ & Submission time and run time of task $j$\\
$\underline t_j$, $\bar t_j$ & Earliest and latest start period of task $j$\\
$D_j$ & Delay limit of task $j$ (periods)\\
$n_j$ & Number of periods spanned by the run of task $j$\\
$a_{j,i}$ & Share of the $i$th period of its run in which task $j$ runs\\
$r_j^c$ & Units of resource $c$ requested by task $j$\\
$R_t^{0,c}$, $\overline R^c$ & Units of resource $c$ used by the running tasks in period $t$; capacity limit of resource $c$\\
$\rho$, $p^c$ & Power usage effectiveness (PUE); power per unit of resource $c$ \textit{above} idle (W)\\
$P_j$ & Facility power added by task $j$ when running (kW)\\
$C_{j,t}$, $\hat C_{j,t}$ & Real-time and expected energy costs of starting task $j$ in period $t$ \\
\midrule
$x_{j,t}$ &     1 if task $j$ starts at the beginning of period $t$, else 0\\
$t_j^\ast$ &    Start period chosen for task $j$\\
$U_t^c$ &       Units of $c$ committed and reserved in period $t$\\
$\mu_t^c$ &     Shadow price of resource $c$ in period $t$ (\$/unit)\\
\bottomrule
\end{tabular}
\end{table}
\setlength{\textfloatsep}{5pt}

\subsection{Problem setting and assumptions} 
\label{sec:model.A}
We take the perspective of an AIDC operator who receives computing tasks and aims to minimize its energy cost as a price taker. The AIDC is exposed to the prices of a two-settlement wholesale market, as in the US or China, either directly or through a retailer that passes these prices on. The day-ahead and real-time prices over period $t$ are denoted by $\lambda_t^{\DA}$ and $\lambda_t^{\RT}$, respectively. The operator schedules tasks in operation periods $t \in \Tset$ of length $\Delta t$, the interval at which it makes decisions, which ranges from seconds to minutes. On-site generations and the adjustment of cooling are not considered.
Tasks submitted before the operating day are already running; together with the idle power, they form the day-ahead schedule, settled at $\lambda_t^{\DA}$. We focus on the tasks submitted during the operating day, which need to schedule in real time. Their energy is a deviation from day-ahead schedule and is settled at $\lambda_t^{\RT}$.

Tasks can be submitted at any second, and every task should be completed. Task $j$ is submitted at $s_j$, runs for $\ell_j$ (spanning $n_j=\lceil\ell_j/\Delta t\rceil$ periods), and requests $r_j^G$ GPUs and $r_j^C$ vCPUs. It can start from period $\underline t_j$, the first period that begins at or after $s_j$, and may wait up to $D_j$ further periods.
Tasks are divided into two categories: 1) High-priority (HP) tasks, such as online inference, are inflexible and start without delay ($D_j=0$); 2) Flexible tasks, such as model training and batch inference, can wait up to $D_j$ periods. Once a task is scheduled, its start cannot be postponed: a user who has been told that a task will finish within the next hour would not accept it finishing two hours later because of the operator's scheduling.

\subsection{Optimization problem formulation} 
\label{sec:model.B}
The AIDC operator minimizes the real-time cost of the tasks submitted during the day over a horizon of $T$ periods. Given all tasks and prices, the offline schedule solves the following mixed-integer linear programming (MILP) problem:
\begin{equation}
    \min_{x}\ \sum_{j\in\J}\sum_{t\in\Sset_j}C_{j,t}\,x_{j,t}
\label{eq:obj}
\end{equation}
subject to
\begin{equation}
    \sum_{t\in\Sset_j}x_{j,t}=1,\quad x_{j,t}\in\{0,1\},\quad\forall j\in\J,
\label{eq:one}
\end{equation}
\begin{subequations}
\label{eq:cap}
\begin{gather}
    R_t^{0,G}+\sum_{j\in\J}r_j^{G}\!\!\sum_{\substack{k\in\Sset_j\\ t-n_j<k\le t}}\!\!x_{j,k}\le\overline R^{G},\quad\forall t\in\Tset,
    \label{eq:cap-gpu}\\
    R_t^{0,C}+\sum_{j\in\J}r_j^{C}\!\!\sum_{\substack{k\in\Sset_j\\ t-n_j<k\le t}}\!\!x_{j,k}\le\overline R^{C},\quad\forall t\in\Tset.
    \label{eq:cap-cpu}
\end{gather}
\end{subequations}
The binary $x_{j,t}$ equals 1 if task $j$ starts at the beginning of period $t$, and $C_{j,t}$ is the real-time energy cost of that start, which is a scalar constant. Constraint~\eqref{eq:one} gives each task exactly one start period. Constraints~\eqref{eq:cap-gpu}--\eqref{eq:cap-cpu} keep the GPU and vCPU use within the resource capacities.
$R_t^{0,G}$ and $R_t^{0,C}$ are the GPUs and vCPUs used in period $t$ by tasks already scheduled in prior days (i.e., outside $\J$).

\textit{Start periods:} Task $j$ can start in the following periods:
\begin{equation}
\begin{aligned}
    &\Sset_j = \{\underline t_j,\ldots,\bar t_j\},
\label{eq:startset}
\end{aligned}
\end{equation}
where $\underline t_j=\lceil s_j/\Delta t\rceil+1$ is the first period starting at or after $s_j$, and $\bar t_j=\min\{\underline t_j+D_j,T\}$ is the latest allowed start period.

\textit{Power and cost:} 
With a fixed PUE $\rho$, task $j$ adds a constant power $P_j$ while it runs, and starting it in period $t$ costs
\vspace{-0.28em}
\begin{equation}
\begin{aligned}
    &C_{j,t}=\Delta t\, P_j\sum_{i=1}^{n_j}a_{j,i}\,\lambda_{t+i-1}^{\RT},\quad
    P_j=\frac{\rho}{1000}\bigl(p^G r_j^G+p^C r_j^C\bigr),\\
    &a_{j,i}=\min\{1,\ \ell_j/\Delta t-i+1\},\quad i=1,\ldots,n_j,
\end{aligned}
\label{eq:cost}
\end{equation}
where $p^G$ and $p^C$ are the power per GPU and per vCPU \textit{above} idle. A task that starts in period $t$ fills its first $n_j-1$ periods and a share $a_{j,n_j} \in [0,1]$ of its last. This formulation keeps the CPU and GPU well used when $\Delta t$ is large. A smaller $\Delta t$ gives each task more start periods, and hence more binary variables, so the MILP may not be solved within one period.

Overall, problem \eqref{eq:obj}--\eqref{eq:cap} minimizes the real-time energy cost. Since $C_{j,t}$ is a constant, the problem is linear in $x$ and is a MILP, with integrality only through $x_{j,t}\in\{0,1\}$. 

\section{Solution Method}
\label{sec:method}
The formulation in \eqref{eq:obj}--\eqref{eq:cap} cannot be applied directly during operation for three reasons. First, tasks arrive one at a time, so in period $t$ only the tasks submitted so far are known, and the global optimum cannot be obtained without the complete task information. Second, the AIDC operator's decision every $\Delta t$ (by solving a MILP over the waiting tasks) may not be finished within a period of seconds, but solving it less often, e.g. every 15~min, makes arriving tasks wait for the next solve. Third, the real-time price is known only for the current settlement interval, so later prices must be forecast. 

\begin{figure}[!t]
\vspace{-1.28em}
    \centering
    \includegraphics[width=0.8\columnwidth]{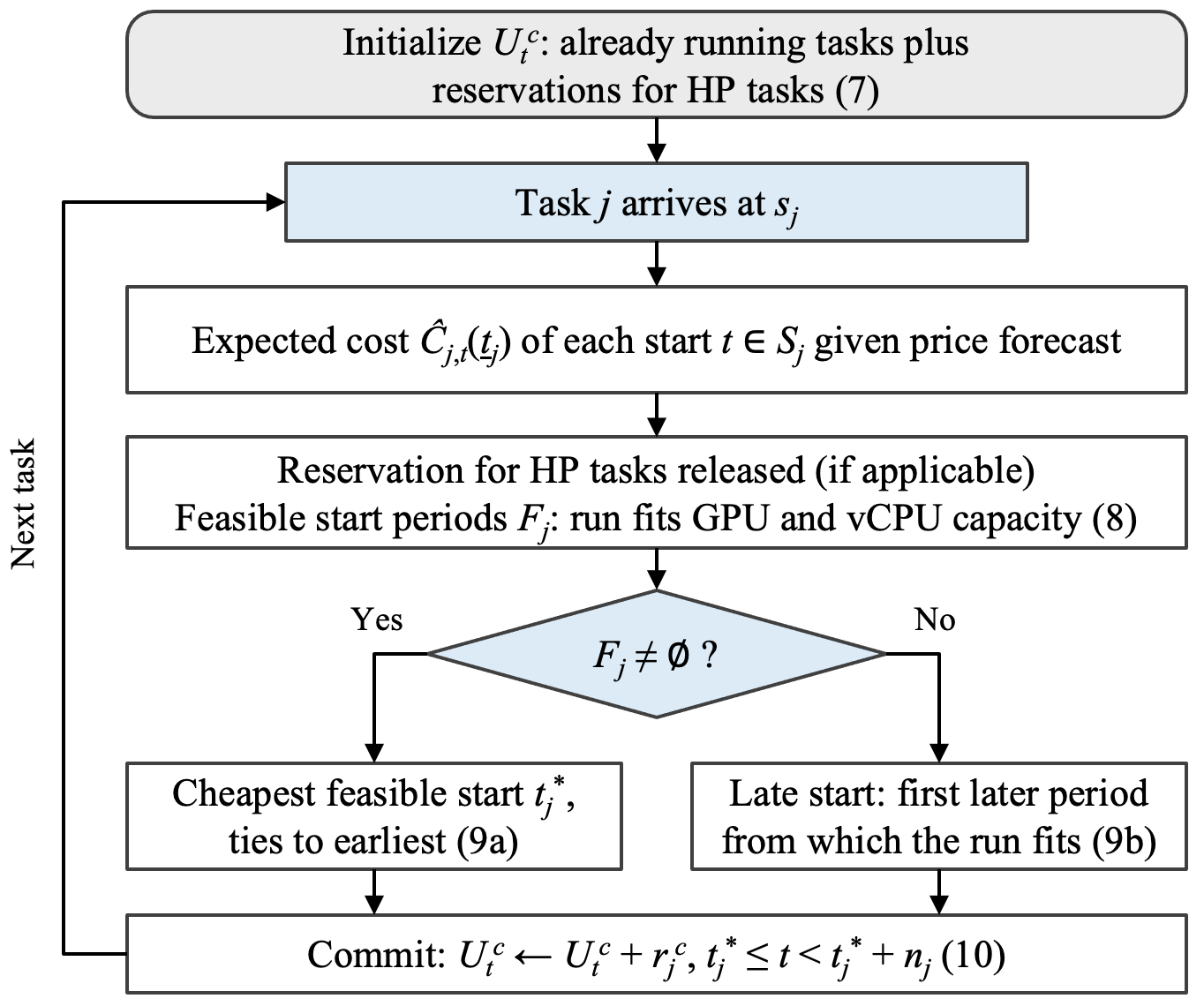}
    \vspace{-1.28em}
    \caption{Workflow of the proposed solver-free online scheduling rule.}
    \label{fig:method}
\vspace{-0.58em}
\end{figure}

To address these limits, we propose a solver-free online rule that assigns each task a start period upon arrival while respecting GPU and CPU capacity constraints. The workflow of the proposed method is illustrated in Fig.~\ref{fig:method}.

\subsection{Online Scheduling Rule}
\label{sec:online}
This section presents the proposed online rule, which schedules the tasks of problem~\eqref{eq:obj}--\eqref{eq:cap} as they arrive. Each task is scheduled once, at its submission time $s_j$, in order of submission.

\textit{Expected cost:} A task submitted at $s_j$ can run from period $\underline t_j$ onward. With the forecast of real-time prices denoted by $\hat\lambda_{t}$, for a task $j$, the expected cost of starting it in period $t$ is
\begin{equation}
\vspace{-0.28em}
\begin{aligned}
    &\hat C_{j,t}=\Delta t \cdot P_j\sum_{i=1}^{n_j}a_{j,i}\,\hat\lambda_{t+i-1}.
\end{aligned}
\label{eq:expcost}
\end{equation}

\textit{Resource capacity:} $U_t^C$ and $U_t^G$ hold the use of GPU and CPU resources committed in period $t$, plus a reservation for upcoming HP tasks,
\begin{subequations}
\label{eq:state}
\begin{gather}
    U_t^{G}=R_t^{0,G}+\sum_{\substack{j\in\J^{\HP}\\ \underline t_j\le t<\underline t_j+n_j}}r_j^{G},\quad\forall t\in\Tset,
    \label{eq:state-gpu}\\
    U_t^{C}=R_t^{0,C}+\sum_{\substack{j\in\J^{\HP}\\ \underline t_j\le t<\underline t_j+n_j}}r_j^{C},\quad\forall t\in\Tset.
    \label{eq:state-cpu}
\end{gather}
\end{subequations}

\textit{Decision for task $j$:} An HP task first releases its own reservation. The feasible start periods $k\in\Sset_j$ are then those from which the task fits, for both GPUs and CPUs, in every period $t$ of its run:
\vspace{-0.28em}
\begin{equation}
\begin{aligned}
    &U_t^G\leftarrow U_t^G-r_j^G,\ \ U_t^C\leftarrow U_t^C-r_j^C,\\
    &\hphantom{U_t^G\leftarrow{}}\underline t_j\le t<\underline t_j+n_j,\quad\text{if } j\in\J^{\HP},\\
    &\F_j=\bigl\{k\in\Sset_j:\ U_t^G+r_j^G\le\overline R^G,\ U_t^C+r_j^C\le\overline R^C,\\
    &\hphantom{\F_j=\bigl\{k\in\Sset_j:\ }\forall t,\ k\le t<k+n_j\bigr\}.
\end{aligned}
\label{eq:feasible}
\end{equation}
If $\F_j\neq\varnothing$, the task starts in its cheapest feasible period, with ties going to the earliest~\eqref{eq:choice}. Otherwise it starts late, in the first period after $\bar t_j$ from which it fits for its whole run~\eqref{eq:late}:
\vspace{-0.48em}
\begin{subequations}
\label{eq:start}
\begin{align}
    t_j^\ast&=\min\Bigl\{\arg\min_{t\in\F_j}\hat C_{j,t}(\underline t_j)\Bigr\},
    \label{eq:choice}\\
    t_j^\ast&=\min\bigl\{k>\bar t_j:\ U_t^G+r_j^G\le\overline R^G,\ U_t^C+r_j^C\le\overline R^C,\notag\\
    &\hphantom{{}=\min\bigl\{k>\bar t_j:\ }\forall t,\ k\le t<k+n_j\bigr\}.
    \label{eq:late}
\end{align}
\end{subequations}
If no such period exists within the horizon, the task is not started. The start is then confirmed:
\vspace{-0.48em}
\begin{equation}
    U_t^G\leftarrow U_t^G+r_j^G,\ \ U_t^C\leftarrow U_t^C+r_j^C,\ \ t_j^\ast\le t<t_j^\ast+n_j.
    \label{eq:commit}
\vspace{-0.28em}
\end{equation}
Each decision evaluates \eqref{eq:expcost} and \eqref{eq:feasible} only over the task's allowed start periods, without calling a solver.

\textit{Feasibility:} Each task receives one start from \eqref{eq:start}, and \eqref{eq:commit} fixes it, so \eqref{eq:one} holds. Condition \eqref{eq:feasible} reflects \eqref{eq:cap} for task $j$ with all earlier starts fixed in $U_t^c$, and \eqref{eq:commit} adds only starts that fit; hence $U_t^c\le\overline R^c$ holds throughout. Each HP task's capacity is reserved in advance, so releasing and recommitting it leaves $U_t^c$ unchanged if task forecast is perfect. Hence every HP task starts on time, provided the running tasks and all HP tasks together fit within capacity. The delay limit holds because $\F_j\subseteq\Sset_j$, and is exceeded only by a late start~\eqref{eq:late}.

\textit{Remark (dual-decomposition view):} Relaxing \eqref{eq:cap} with shadow prices $\mu_t^c\ge0$ splits \eqref{eq:obj}--\eqref{eq:cap} into one problem per task:
\vspace{-0.58em}
\begin{equation}
    t_j(\mu)\in\arg\min_{t\in\Sset_j}\Bigl[\,C_{j,t}+\sum_{c\in\{G,C\}}\ \sum_{k=t}^{t+n_j-1}\mu_{k}^c\,r_j^c\Bigr].
    \label{eq:dual}
\vspace{-0.28em}
\end{equation}
The first term is the energy cost; the second is the cost of the capacity that the start occupies. Dual decomposition updates $\mu_t^c$ by subgradient steps until the starts fit~\cite{boyd2008}; this needs all tasks and many iterations. The online rule instead sets the prices from the committed state:
\vspace{-0.38em}
\begin{equation}
    \mu_t^c=\begin{cases}0,& U_t^c+r_j^c\le\overline R^c,\\ +\infty,&\text{otherwise.}\end{cases}
    \label{eq:mu}
\vspace{-0.28em}
\end{equation}
With \eqref{eq:mu},
\eqref{eq:dual} becomes \eqref{eq:feasible}--\eqref{eq:choice}: start periods whose run meets a period without room for the task are excluded, and the cheapest remaining one is taken. The reservation \eqref{eq:state} prices HP capacity at $+\infty$ for flexible tasks. If capacity is slack at the cheapest starts, $\mu_t^c=0$ is optimal and \eqref{eq:dual} reduces to the cheapest start; with all real-time prices known, the rule then solves \eqref{eq:obj}--\eqref{eq:cap}. If capacity binds, the optimal $\mu_t^c$ is finite and assigns a congested period to the task that gains most from it, whereas \eqref{eq:mu} assigns it to the first arrival. 
\vspace{-0.28em}

\section{Case Study}
\vspace{-0.48em}
\label{sec:results}
\subsection{Case Setup}
\vspace{-0.18em}
We consider the operation of a AIDC cluster located in California, US. It has 3,456 A100-SXM4-80GB GPUs and 55,296 vCPUs. Each GPU draws 50~W when idle and 400~W at full load, and each vCPU 1~W and 5~W. With a PUE of $\rho=1.30$, the facility power ranges from 297~kW with all servers idle to 2,157~kW at full load.

\begin{figure}[!t]
\vspace{-0.00em}
    \centering
    \includegraphics[width=0.8\columnwidth]{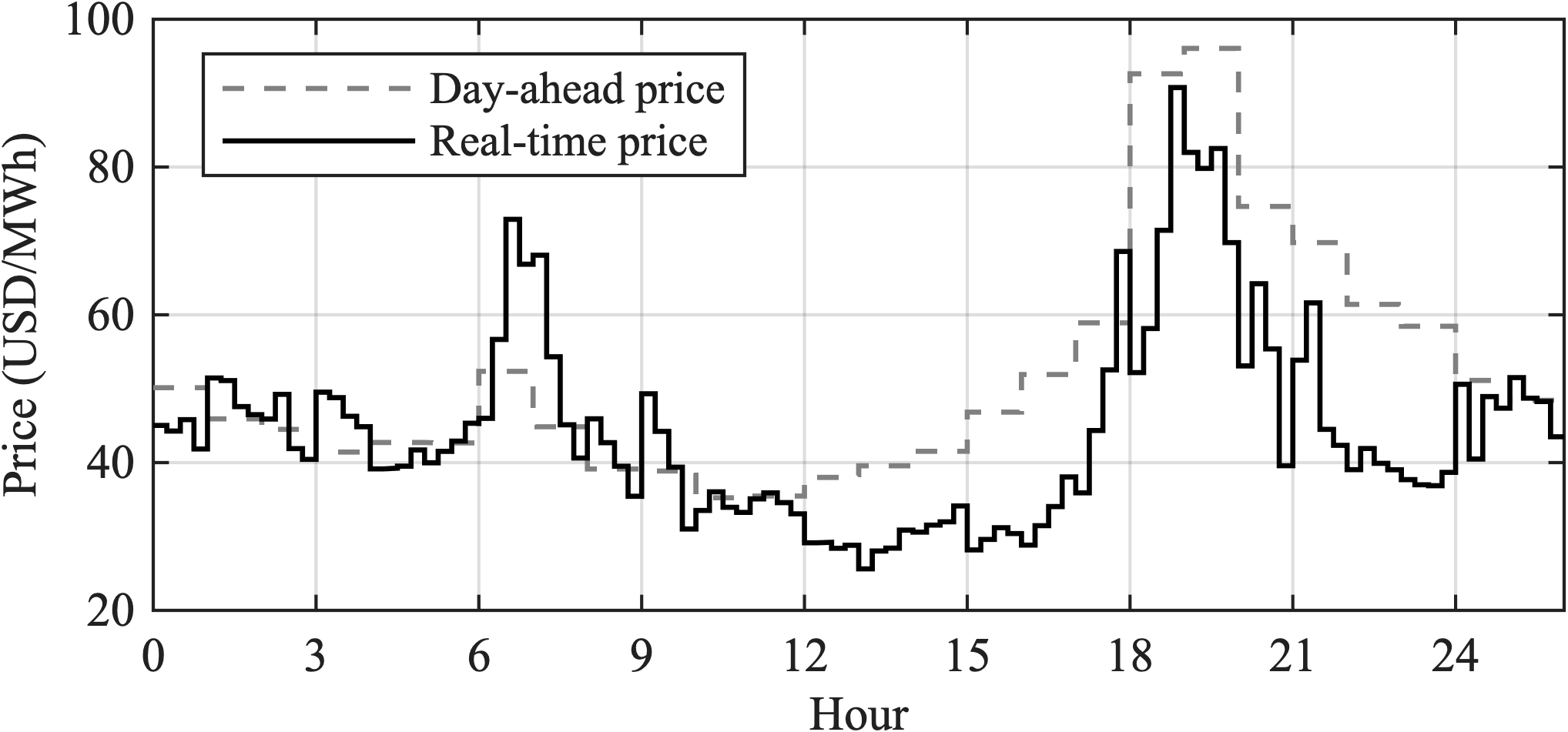}
    \vspace{-0.88em}
    \caption{Day-ahead and real-time prices of CAISO NP15 on August 21, 2026.}
    \label{fig:price}
\end{figure}

We consider that the AIDC is exposed to CAISO wholesale prices, including two parts:
The day-ahead part, for the idle power and the tasks already running from the previous day, and the real-time part, for the tasks submitted during the operating day. We focus on the real-time part.
We use the NP15 locational marginal prices of August 21, 2026 (Fig.~\ref{fig:price}).

The tasks come from the job set of the reference~\cite{duan2026gfs}, which records the submission time, duration and GPU and CPU requests of each job. We make two changes for our tests. First, we reassign each job to one of three priority classes, HP, Priority and Spot, while keeping its other features. Second, to stress the capacity logic, we move jobs recorded on other GPU types into the considered task set, so that peak demand reaches about 109.2\% of GPU capacity while all tasks can still be completed within the horizon.

\begin{figure}[!t]
\vspace{-0.88em}
    \centering
    \includegraphics[width=0.8\columnwidth]{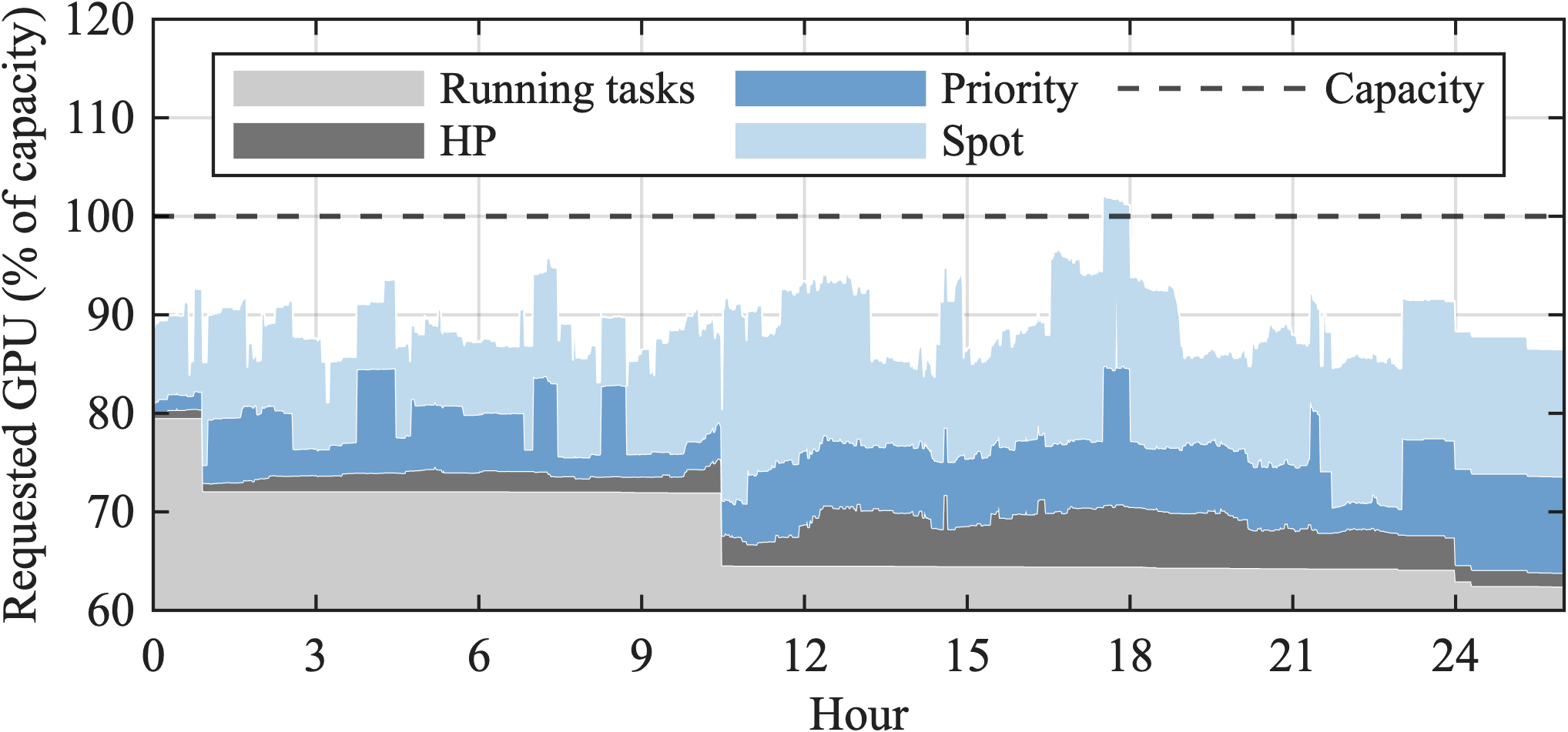}
    \vspace{-0.88em}
    \caption{GPU demand of the tasks if every task starts at its submission: tasks running from yesterday and the HP, Priority and Spot tasks of the day.}
    \label{fig:tasks}
\vspace{-0.00em}
\end{figure}

On the operating day, 1,622 tasks from yesterday are already running at 00:00, and 1,343 new tasks are submitted during the day (350 HP, 311 Priority and 682 Spot), which may arrive at any second, as shown in Fig.~\ref{fig:tasks}.

We compare four cases: 1) Proposed: the solver-free online rule; 2) Benchmark~1 (earliest feasible): starts each task in its earliest feasible period without considering energy cost; 3) Benchmark~2 (rolling MILP): re-solves \eqref{eq:obj}--\eqref{eq:cap} for waiting tasks at each update every 5~min (following multi-timescale dispatch); and 4) Perfect (offline MILP): solves the full-horizon problem \eqref{eq:obj}--\eqref{eq:cap} after the day with all tasks and prices known, providing a perfect-information reference with the global optimum. For a fair comparison of computation time, the MILPs of Benchmark~2 and Perfect are solved by calling the MATLAB \texttt{intlinprog} solver directly, without a modeling layer such as YALMIP, with a relative gap of $10^{-4}$. The simulation settings are summarized in Table~\ref{tab:settings}.

\begin{table}[!t]
\vspace{-0.00em}
\caption{Simulation Settings}
\vspace{-0.88em}
\label{tab:settings}
\centering
\footnotesize
\setlength{\tabcolsep}{5pt}
\begin{tabular}{@{}ll@{\hspace{8pt}}ll@{}}
\toprule
Parameter & Value & Parameter & Value\\
\midrule
Horizon & 26 h & $D_j$, Spot & 120 (120 min)\\
Period length $\Delta t$ & 1 min & Benchmark 2 update & 5 min\\
$D_j$, HP & 0 & Offline MILP gap & $10^{-4}$\\
$D_j$, Priority & 60 (60 min) & Offline time limit & 600 s\\
\bottomrule
\end{tabular}
\vspace{-0.88em}
\end{table}

\begin{table}[!t]
\vspace{-0.28em}
\caption{Energy Cost, Scheduling Performance, and Computation Time}
\vspace{-0.88em}
\label{tab:results}
\centering
\footnotesize
\setlength{\tabcolsep}{6pt}
\begin{tabular}{@{}lrrrr@{}}
\toprule
Method & \shortstack[r]{Flexible-task\\cost (\$)} & \shortstack[r]{Cost above\\offline (\%)} & \shortstack[r]{Late HP\\tasks (\#)} & \shortstack[r]{Time per\\decision (ms)}\\
\midrule
Benchmark 1 & 378.02 & 12.9 & 0 & 0.032\\
Benchmark 2 & 341.27 & 1.9 & 7 & 41.4\\
\rowcolor{gray!15} Proposed & 340.74 & 1.7 & 0 & 0.024\\
Perfect & 334.97 & 0 & 0 & 52,800\\
\bottomrule
\end{tabular}
\end{table}

\subsection{Cost and Computation}
\vspace{-0.00em}
Table~\ref{tab:results} shows the energy cost and scheduling performance of flexible tasks (only Priority and Spot).
Benchmark~1 starts each task in its earliest feasible period. The proposed method instead delays Priority and Spot tasks by 18.9 and 39.1~min on average, shifting them to lower-price periods. This lowers the flexible-task cost by \$37.28 (9.9\%). The proposed method also costs \$0.53 (0.16\%) less than Benchmark~2.
Although this saving is modest, Benchmark~2 has three drawbacks: 1) it takes over 1,000 times longer per decision compared to the proposed method; 2) since it updates only every 5~min, tasks wait 1.2~min longer on average; and 3) with no capacity reserved for HP tasks, seven HP tasks start late, violating their timing requirements. (We did not add penalty in the reported cost.)
Finally, the proposed method costs \$5.77 (1.7\%) more than Perfect. However, Perfect requires complete knowledge of future tasks and prices and takes 52.8~s (nearly 1~min) to solve the full-horizon problem, making it impractical for real-time operation.

\subsection{Capacity and HP Tasks}
\begin{figure}[!t]
\vspace{-0.00em}
    \centering
    \includegraphics[width=0.85\columnwidth]{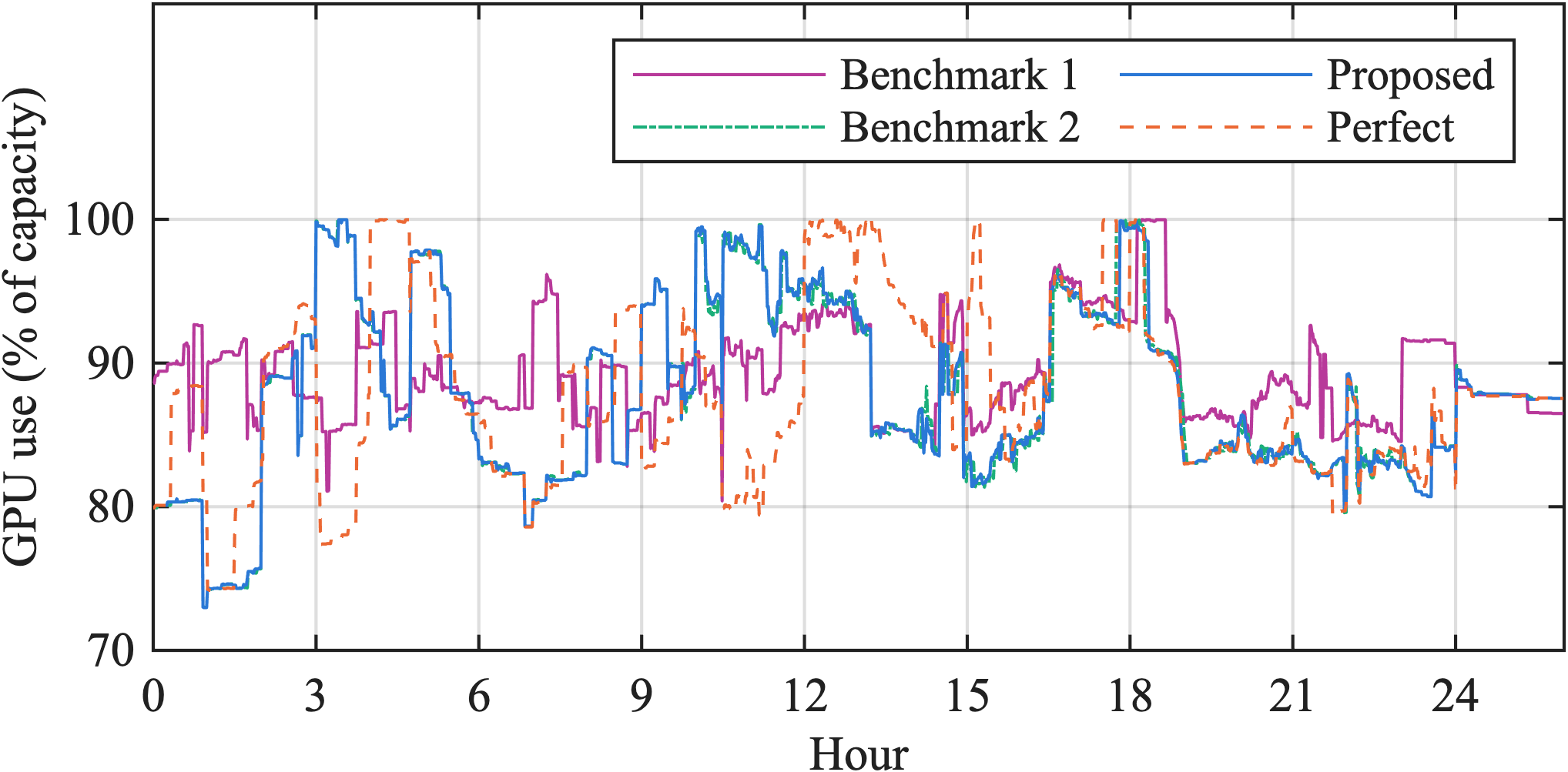}
    \vspace{-0.88em}
    \caption{GPU use per period of the four cases.}
    \label{fig:gpu}
\vspace{-0.58em}
\end{figure}

\begin{figure}[!t]
    \centering
    \includegraphics[width=0.85\columnwidth]{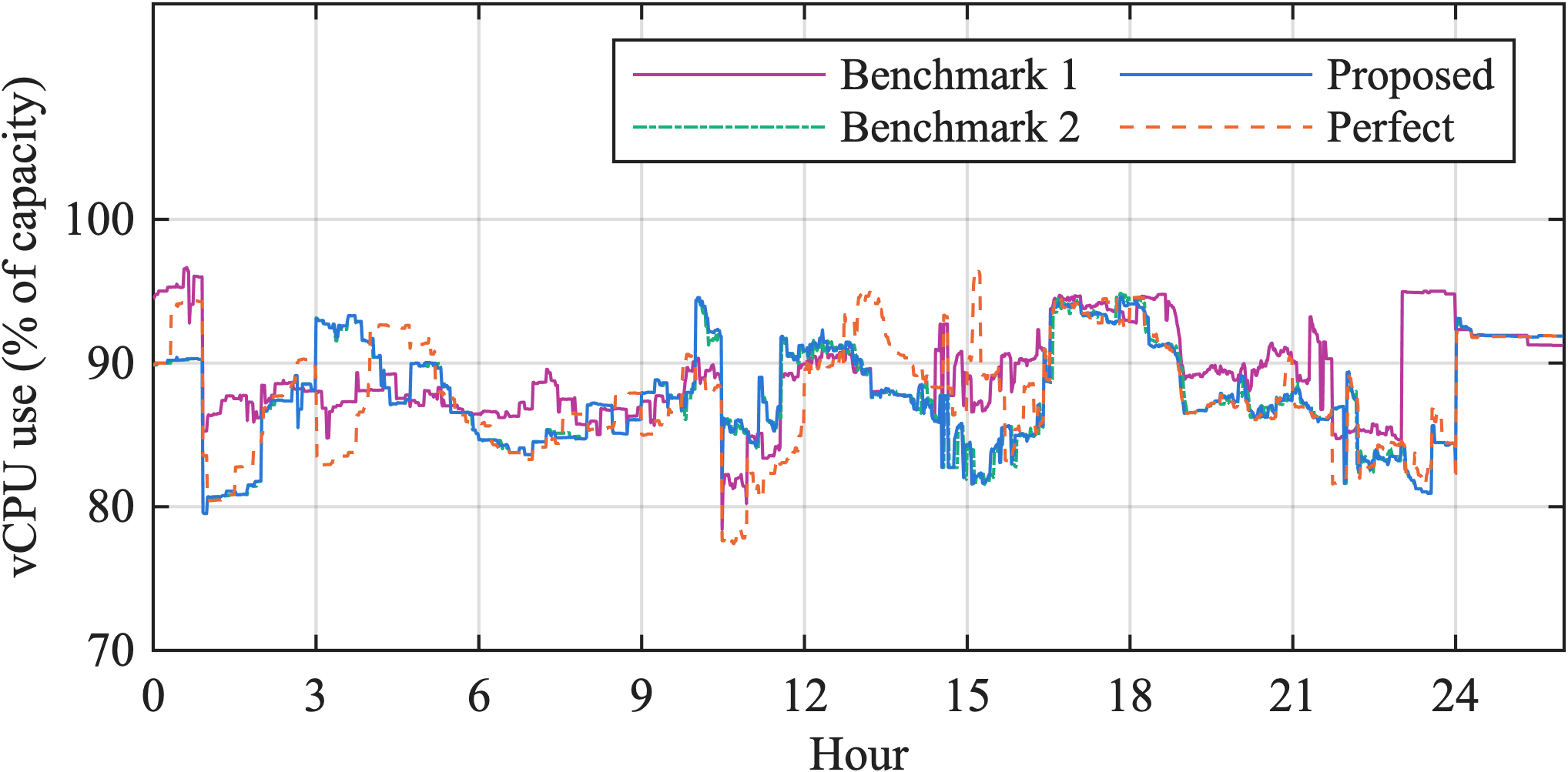}
    \vspace{-0.88em}
    \caption{vCPU use per period of the four cases.}
    \label{fig:cpu}
\vspace{-0.00em}
\end{figure}

\vspace{-0.00em}
GPU capacity is the binding constraint: utilization reaches 100\% around 03:00, 10:00--12:00 and 17:00--19:00 under the proposed method (Fig.~\ref{fig:gpu}), while vCPU utilization remains below 97\% across all methods (Fig.~\ref{fig:cpu}). Both the proposed method and Benchmark~1 reserve capacity for future HP tasks and start all HP tasks on time. However, Benchmark~1 does not consider the impact of electricity prices, leaving GPU load high during the high-price periods around {09:00 and 19:00 }(Fig.~\ref{fig:gpu}), whereas the proposed method combines capacity reservation with price-aware scheduling. Benchmark~2 also uses price forecasts but lacks capacity reservation. Earlier batches can therefore occupy the available GPU capacity, delaying seven HP tasks by 5--32~min.

\subsection{Impact of the Price and Task Forecast}
On the operating day, the day-ahead price follows the shape of the real-time price (Fig.~\ref{fig:price}), so it serves as a reasonable forecast. The proposed method and Benchmark 2 use the same information: the day-ahead price as the price forecast, and perfect knowledge of the HP arrivals.

\begin{figure}[!t]
\vspace{-0.00em}
    \centering
    \includegraphics[width=0.85\columnwidth]{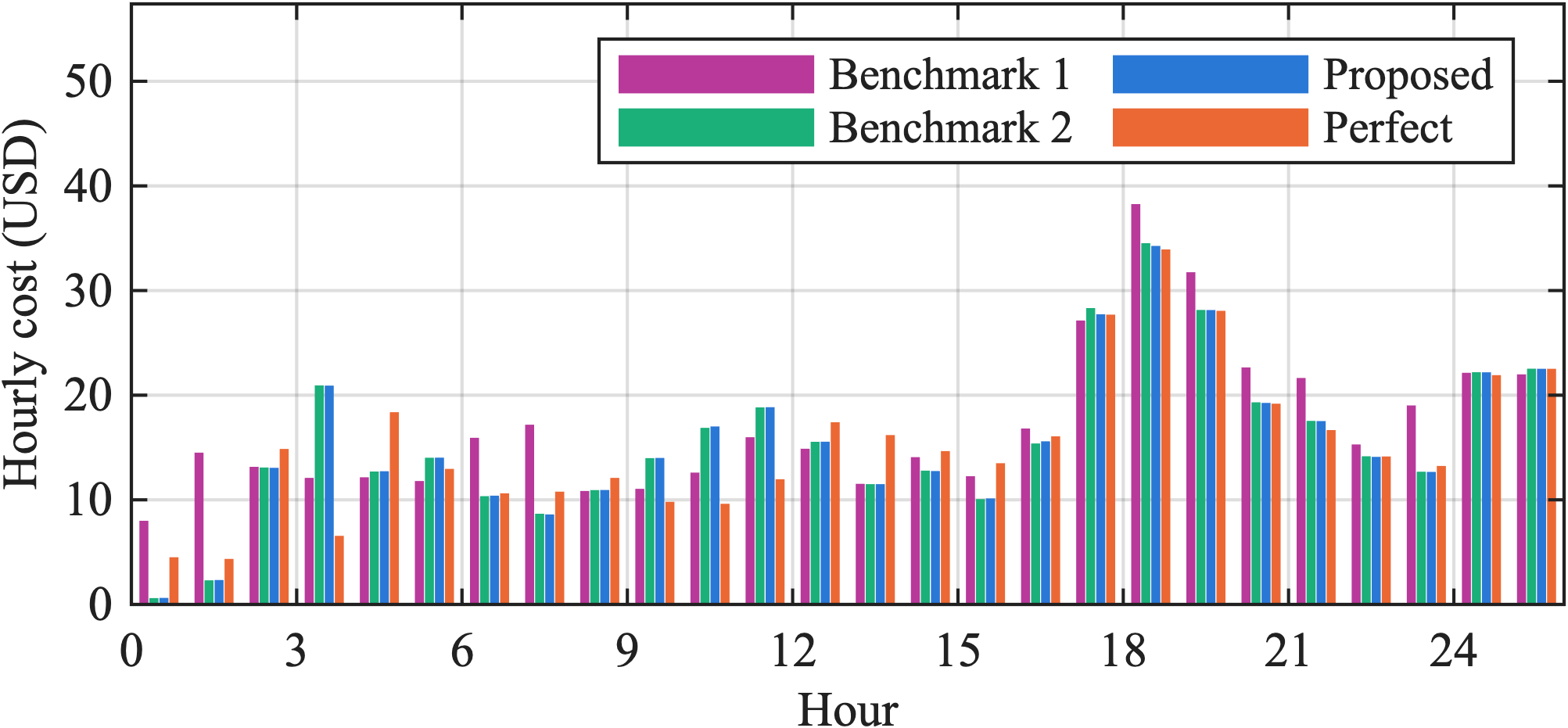}
    \vspace{-0.88em}
    \caption{Hourly cost of the tasks of the day.}
    \label{fig:hourly}
\vspace{-0.88em}
\end{figure}

\begin{figure}[!t]
    \centering
    \includegraphics[width=0.85\columnwidth]{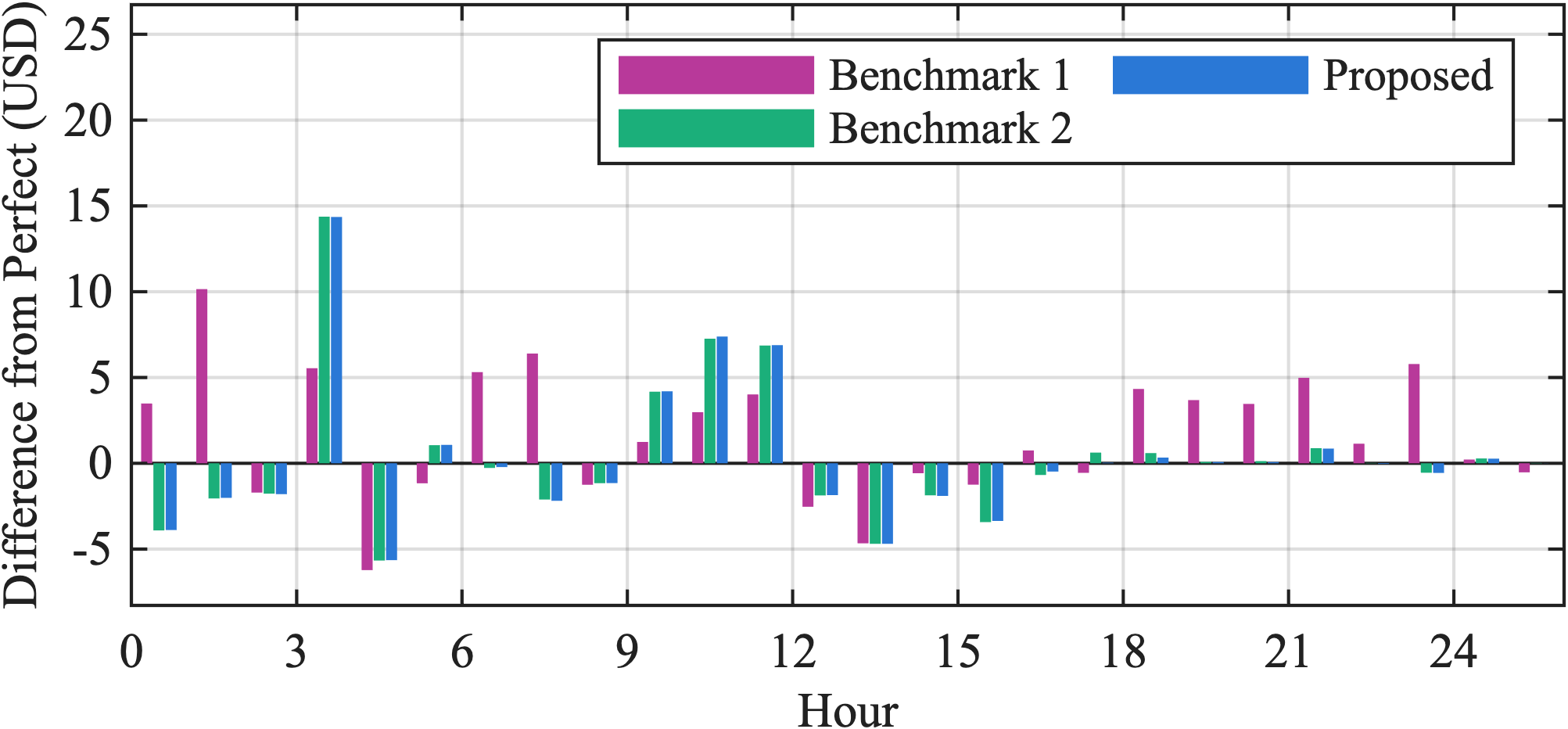}
    \vspace{-0.88em}
    \caption{Difference in hourly cost relative to Perfect.}
    \label{fig:difference}
\end{figure}

With capacity mostly slack, $\mu_t^c=0$ in most periods, so both methods take the cheapest start under the forecast, and their hourly costs differ by at most \$0.6 (Fig.~\ref{fig:hourly}). The gap to Perfect comes from forecast error: between 12:00 and 17:00, the real-time price falls 9--19~\$/MWh below the day-ahead price, so Perfect shifts load from 09:00--12:00 to 12:00--16:00 (Fig.~\ref{fig:difference}). From 17:00, when the forecast matches the price, the hourly costs agree within \$1. This shows the value of better price forecasts.

\section{Conclusion}
\label{sec:conclusion}
In this work, we propose an online solver-free rule that schedules each task of an AIDC once, at submission, in its cheapest capacity-feasible start period under the forecast real-time price, given the day-ahead schedule. 
Under a stressed workload, the proposed method has a 9.9\% lower cost than the earliest-feasible rule (Benchmark~1). Compared with a rolling MILP (Benchmark~2), it achieves a slightly lower cost, starts all HP tasks on time, and needs 0.024~ms per decision against 41.4~ms per update. Its cost is within 1.7\% of the perfect offline optimum.
Future work will address task and market price uncertainty, 
multi-day evaluations, and the theoretical bound of the online algorithm.

\bibliographystyle{IEEEtran}
\bibliography{references}

\end{document}